\documentclass{PoS}

\title{Low energy properties of color-flavor locked superconductors }

\ShortTitle{Low energy properties of color-flavor locked superconductors }

\author{\speaker{Cristina Manuel}\\
        Instituto de Fisica Corpuscluar (CSIC-U. de Val\`encia)\\
        E-mail: \email{cristina.manuel@ific.uv.es}}

\def\bea{\arraycolsep .1em \begin{eqnarray}}
\def\eea{\end{eqnarray}}

\newcommand{\cd}{\! \cdot \!}
\newcommand{\be}{\begin{equation}}
\newcommand{\ee}{\end{equation}}
\newcommand{\ba}{\begin{eqnarray}}
\newcommand{\ea}{\end{eqnarray}}

\def\lsim{\mathrel{\lower4pt\hbox{$\sim$}}\hskip-12.5pt\raise1.6pt\hbox{$<$}\;}
\def\gsim{\mathrel{\lower4pt\hbox{$\sim$}}
\hskip-12.5pt\raise1.6pt\hbox{$>$}\;}

\def\s0#1#2{\mbox{\small{$ \frac{#1}{#2} $}}}
\def\0#1#2{\frac{#1}{#2}}

\abstract{ We discuss some low energy properties of  color-flavor
locked (CFL) superconductors. First, we study how an external
magnetic field affects their Goldstone physics in the chiral limit, stressing that there is a long-range component of the field that penetrates the superconductor.  We note that the most
remarkable effect of the applied field is giving a mass to the charged pions and kaons.
By estimating this effect, we see that for values ${\widetilde
e}{\widetilde B} \sim 2 f_\pi \Delta$, where $\Delta$ is the quark
gap, and $f_\pi$ the pion decay constant, the charged Goldstone bosons
become so heavy, that they turn out to be unstable. The symmetry breaking
pattern is then changed, agreeing with that of the magnetic color-flavor locked (MCFL)
phase, recently proposed in hep-ph/0503162.
Finally, we discuss the physics of the superfluid phonon of the
CFL phase, compare it with that of the phonon of a Bose-Einstein condensate,
and discuss transport phenomena at low temperature. Astrophysical implications of all the
above low energy properties are also commented.}

\FullConference{29th Johns Hopkins Workshop on current problems in particle theory:
strong matter in the heavens\\
                 1-3 August\\
                 Budapest}

\begin{document}

%\maketitle  IS IGNORED %%%%%%%%%%%

%%%%%%%%%%%%%%%%%%%%%%%%%%%%%%%%%%%%%%%%%%%%%%%%%%%%%%%%%%%%%%%%%%%%%%%%%%%%%%%%%%%%%%%%%%%%%%%%%%

\section{Introduction}

%\section{Color superconductivity in an external strong magnetic field}

In this talk I will touch two different aspects of the low energy physics of the color-flavor
locked phase \cite{Alford:1999mk}.
The first part is a continuation of Vivian de la Incera's seminar, on the subject
of how a strong magnetic field influences color superconductivity. It is based on the joint collaboration
with Efrain Ferrer and Vivian de la Incera \cite{Ferrer:2005vd}.
 In the second part, I will discuss transport phenomena
at low temperature of the CFL and BEC superfluids, stressing similarities and differences. This is
work done in collaboration with Antonio Dobado and Felipe Llanes-Estrada \cite{Manuel:2004iv},
and an on-going collaboration
with Arnau Rios. Before entering into detailed discussions, let me explain you
why those are interesting physical problems.

Studying the influence of a magnetic field on quark matter is not only an academic question.
We believe that quark matter may occur in compact stars, either in the core of neutron stars, or in the form of quark stars.
The real fact is that the highest magnetic fields in the Universe occur precisely in compact stars. Pulsars are believed to
be neutron stars, and they stand  magnetic fields in their surfaces in the range $B \sim 10^{12} - 10^{14}$ G.
Magnetars are a different kind of compact stars that stand higher magnetic fields, $B \sim 10^{14} - 10^{15}$ G,
while the value $B \sim 10^{16}$ G is yet not discarded. There is however un upper theoretical limit to
the magnetic field that a compact star may stand, arising after comparing the magnetic and gravitational
energy, given as \cite{Lai:2000at}

\begin{equation}
B_{\rm max} \sim 1.4 \times 10^{18} \left( \frac{M}{M_{\bigodot}} \right) \left( \frac{10 \,{\rm km}}{R} \right)^2
G
\end{equation}
where $M$ and $R$ refer to the mass and radius of the star, respectively, and $M_{\bigodot}$ is the solar mass.
If self-bound quark stars exist, this upper limit may go higher, though.

As discovered in \cite{Ferrer:2005vd},
not only an applied magnetic field to the superconductor affects the quark gaps, thus affecting the equation of
state of CFL quark matter.
It also affects its low energy properties, as we will further discuss.

On the other hand, let me also insist on the relevance of transport phenomena
in astrophysics. While with the equation of state one can  determine
the mass and radius of a star, with the transport coefficients one
can study its cooling, and its vibrational and rotational properties.
Thus, they are essential for detecting signatures of quark matter
in compact stars in any of its possible phases.

It has been established that the viscosities put stringent tests to astrophysical
models for very rapidly rotating stars, such as for millisecond pulsars.
This is based on the existence of r(otational)-mode instabilities in all
relativistic rotating stars \cite{Andersson:1997xt}, which are only
suppressed by sufficiently large viscosities.
So the viscosities will allow one to discard unrealistic models for
millisecond pulsars. There has not been many efforts in the literature to
study transport phenomena in color superconducting quark matter, with very
few exceptions \cite{Madsen:1999ci,Litim:2001je,Shovkovy:2002kv,Manuel:2004iv}.

\section{Effects of a magnetic field in the low energy physics of the CFL phase}

For all the considerations to follow, we will neglect the effect of quark masses. In the absence of a magnetic field,
three-flavor massless quark matter at high baryonic density is in the CFL phase \cite{Alford:1999mk}. Then the diquark condensates lock the color and flavor
transformations, breaking both. The symmetry breaking pattern  in the CFL phase is
\begin{equation}
 SU(3)_C \times SU(3)_L \times SU(3)_R \times U(1)_B
\rightarrow SU(3)_{C+L+R} \ .
\end{equation}
There are only nine Goldstone bosons that survive to the
Anderson-Higgs mechanism. One is a singlet, scalar mode,
associated to the breaking of the baryonic symmetry, and the
remaining octet is associated to the axial $SU(3)_A$ group, just
like the octet of mesons in vacuum. At sufficiently high density, the anomaly is
suppressed, and then one can as well consider the spontaneous breaking of an approximated
$U(1)_A$ symmetry, and  an additional pseudo
Goldstone boson. We will ignore this effect, though.

Once electromagnetic effects are considered, the flavor symmetries
of QCD are reduced, as only the $d$ and $s$ quarks have equal
electromagnetic charges, $q= -e/3$, while the $u$ quark has
electromagnetic charge, $q=2e/3$. However, because the
electromagnetic structure constant $\alpha_{\rm e.m.}$ is so
small, this effect is considered to be really tiny, a small
perturbation, and one talks about approximated good flavor symmetries.

In the CFL phase there is a linear combination of the photon and a gluon that
remains massless. The CFL diquarks are invariant under a ${\widetilde U(1)}_{\rm e.m.}$
group, generated in flavor-color space by $ {\widetilde Q} = Q \times 1 - 1 \times Q$,
where $Q$ is the electromagnetic charge generator. Then quarks of different flavors
and colors all get integral value charges, given in terms of the charge of the electron
${\tilde e} = e \cos{\theta}$, where $\theta$ is the mixing angle.

 The existence of this
``rotated" electromagnetism implies, among other things, that an external magnetic field
to the color superconductor will be able to penetrate it in the form of a ``rotated" magnetic
field, and this affects the pairing phenomena \cite{Ferrer:2005vd}. Furthermore,
in the presence of a strong magnetic field one cannot consider the effects of electromagnetism as a small perturbation.
Flavors symmetries are reduced, as both $U(1)_{\rm e.m.}$ and  ${\widetilde U(1)}_{\rm e.m.}$ distinguish quark flavors.
For sufficiently strong magnetic fields, quark matter is in the so-called {\it magnetic} CFL (MCFL) phase \cite{Ferrer:2005vd}.
In the MCFL phase
the symmetry breaking pattern is
\begin{equation}
SU(3)_C \times SU(2)_L \times SU(2)_R \times U(1)^{(1)}_A\times U(1)_B \times U(1)_{\rm e.m.}
\rightarrow SU(2)_{C+L+R} \times {\widetilde U(1)}_{\rm e.m.} \ .
\end{equation}
Here the symmetry group $U(1)^{(1)}_A$ is related to a current
which is an anomaly free linear combination of $u,d$ and $s$ axial
currents, and such that  $U(1)^{(1)}_A \subset SU(3)_A$. The locked $SU(2)$ group
corresponds to the maximal unbroken symmetry, such that it maximizes the condensation
energy. The counting of broken generators, after taking into account the
Anderson-Higgs mechanism, tells us that there are only five Goldstone
bosons. As in the CFL case, one is associated to the breaking of
the baryon symmetry; three Goldstone bosons are associated to the
breaking of $SU(2)_A$, and another one associated to the breaking
of  $U(1)^{(1)}_A$. As before, if the effects of the anomaly could
be neglected, there would be another pseudo Goldstone
boson associated to the $U(1)_A$ symmetry.

An applied strong  magnetic field, apart from modifying the value
of some fermionic gaps, also affects the low energy properties of
the color superconductor. There is a reduction in the number of
Goldstone bosons, from nine to five. What happened to these four
light particles of the CFL phase?

There is another question one may like to address. For which
values of the magnetic field  it is more appropriate to
say that quark matter is  in the CFL or in the MCFL phase? To
answer properly this question, one should solve the whole set of
gap equations for all values of the magnetic field, a rather
difficult task that requires a numerical treatment.

There is a fast way to answer to the above questions, which may
allows us to give estimates for the transition values of magnetic field.
We will study how an external magnetic field affects the  effective
field theory of the low energy degrees of freedom of the CFL
phase.

\subsection{Effective field theory for the Goldstone bosons of the CFL phase in a magnetic field}

We review here how to construct the effective field theory for the light degrees
of freedom of the CFL phase \cite{Casalbuoni:1999wu}. First
 one has to  single out the phases of the diquark
condensates
\begin{equation}
X^{ia} \sim \epsilon^{ijk} \epsilon^{abc} \langle \psi^{bj}_L \psi^{ck}_L \rangle^* \ , \qquad
Y^{ia} \sim \epsilon^{ijk} \epsilon^{abc} \langle \psi^{bj}_R \psi^{ck}_R \rangle^*
\end{equation}
where $a,b,c$ denote flavor indices, $i,j,k$ denote color indices, and $L/R$ denote left/right
chirality, respectively. Under an $SU(3)_C \times SU(3)_L \times SU(3)_R$ rotation, the above fields transform as
\begin{equation}
X \rightarrow U_L X U_C^\dagger \ , \qquad Y \rightarrow U_R Y U_C^\dagger \ .
\end{equation}
The combination defined by
\begin{equation}
\Sigma = X Y^\dagger
\end{equation}
is a color singlet, which under $SU(3)_C \times SU(3)_L \times SU(3)_R$ transforms as
\begin{equation}
\Sigma \rightarrow U_L \Sigma U_R^\dagger \ .
\end{equation}

One parametrizes the unitary matrix $\Sigma$ as
\begin{equation}
 \Sigma =  \exp \left(i\,{\Phi\over f_\pi} \right) \ , \qquad
\Phi = \phi^A T^A \ ,
\end{equation}
where $T^A$ are the $SU(3)$ generators,
defining the Goldstone fields as in vacuum
\begin{equation}
\Phi =  \,\left(
\begin{array}{ccc}
 \pi_0 + {1\over \sqrt 3}  \eta_8
  & \sqrt{2} \pi^+ &\sqrt{2} K^+ \\
\sqrt{2} \pi^- & - \pi_0 + {1\over \sqrt 3} \eta_8 & \sqrt{2} K^0\\
\sqrt{2} K^- & \sqrt{2} \bar K^0 &  {-2\over \sqrt 3 }  \eta_8 \\
\end{array}\right) \ .
\end{equation}

For energy scales $E\ll 2 \Delta$, where $\Delta$ is the quark gap, the Lagrangian in
the chiral limit for the Goldstone fields is pretty much the same
as in vacuum, with the only exception that temporal and spatial derivatives
are not related, due to the lack of Lorentz symmetry at finite density
\begin{equation}
\label{chiral-lag}
{\cal L}   = \frac{f_\pi^2}{4} \left(
{\rm Tr} \left( \partial_0 \Sigma
\partial_0 \Sigma^\dagger \right) -
v^2_\pi {\rm Tr} \left( \partial_i \Sigma
\partial_i \Sigma^\dagger \right) \right) \ .
\end{equation}

For asymptotic large values of the chemical potential $\mu$, the parameters appearing in (\ref{chiral-lag}) can
be computed, and read \cite{Son:2000cm}
\begin{equation}
\label{fpi}
f_\pi^2 = {21 - 8 \ln{ 2} \over 18} \, {\mu^2\over 2 \pi^2} \ ,
\qquad v_\pi = \frac{1}{\sqrt{3}} \ .
\end{equation}

We have ignored in all the above considerations the Goldstone boson of baryon symmetry breaking,
which will deserve a separated treatment in Section \ref{transport}.

Electromagnetic interactions can be introduced taking into
account, as commented in the previous section, that they do not
respect the whole $SU(3)_L \times SU(3)_R$ symmetry. To see which
are the new allowed terms in the low energy effective field theory
one proceeds as follows. The quark charge matrix $Q$ represents an
explicit breaking term of the $SU(3)_L \times SU(3)_R$ symmetry in
the QCD Lagrangian. In order to see its effects in the low energy
physics, one treats $Q$ as a spurion field, whose vacuum
expectation value is the responsible of the symmetry breaking. To
restore the symmetry one has to introduce left and right charge
matrices, which transform as
\begin{equation}
Q_L \rightarrow U_L Q_L U_L^\dagger \ , \qquad Q_R \rightarrow U_R Q_R U_R^\dagger \ ,
\end{equation}
under a $SU(3)_L \times SU(3)_R$ transformation.

Then, the  derivatives in Eq.~(\ref{chiral-lag}) should become covariant derivatives
\begin{equation}
\partial_\mu \Sigma \rightarrow {\widetilde D}_\mu \Sigma = \partial_\mu \Sigma - i {\tilde e} Q_L {\tilde A}_\mu \Sigma + i {\tilde e} \Sigma Q_R {\tilde A}_\mu \ .
\end{equation}
Using simple  considerations one can see that also a term of the sort
${\rm Tr} \left( Q_L \Sigma Q_R \Sigma^\dagger \right)$
is  allowed by the symmetries in the Lagrangian.
Thus, it
reads \cite{Litim:2001mv}
\begin{eqnarray}
\label{chiralL+ e.m.}
{\cal L}     &=&
\frac{\widetilde{\epsilon}}{2}\,
{\bf {\widetilde E}}^2
- \frac {1}{2 \widetilde \lambda}\,
{\bf {\widetilde B}}^2
\nonumber \\ & &
+ \frac{f_\pi^2}{4}
\left[ {\rm Tr} \left({\widetilde D}_0 \Sigma
                      {\widetilde D}_0 \Sigma^\dagger \right)
 - v^2_\pi {\rm Tr} \left( {\widetilde D}_i \Sigma
                           {\widetilde D}_i \Sigma^\dagger \right)
\right]
+  C\, {\rm Tr}
  \left( Q \Sigma Q \Sigma^\dagger \right)
 \ ,
\end{eqnarray}
Here, again, we have taken into account that spatial and temporal components of ${\widetilde F}_{\mu \nu}$ go separately.
The value of ${\widetilde{\epsilon}} \approx 1 + \frac{{2\,}}{9 \pi^2}\frac{\widetilde e^2\mu^2}{\Delta^2}$,  and $\widetilde \lambda \approx 1$
are obtained after integrating out the high energy fermionic modes \cite{Litim:2001mv}, and represent an effective dielectric constant and magnetic susceptibility,
respectively.

If one uses the value of $Q =e (2/3,-1/3,-1/3)$, one easily recognizes
that the last term of (\ref{chiralL+ e.m.}) is a mass term for the charged Goldstone
bosons, plus additional meson contact interactions. More
particularly, one finds
\begin{equation}
M^2_{\pi^\pm} = M^2_{K^\pm} = \frac{ 2 {\tilde e}^2 C}{f^2_\pi}  \ .
\end{equation}
The value of $C$ could be in principle computed at high baryonic density \cite{Manuel:2001xt}. However, it
requires to evaluate a complicated three-loop diagram.
An estimate of the diagram \cite{Manuel:2001xt}, together with dimensional analysis, suggests that $M^2_{\pi^\pm} = M^2_{K^\pm} \sim {\tilde e}^2 \Delta^2$.

In writing Eq.(\ref{chiralL+ e.m.}) one assumes a particular power counting. The chiral Lagrangian is written as an expansion
in momenta, and here only the lowest order terms have been kept. With this counting, one assumes $p^2/f_\pi^2 \sim {\tilde e}^2$.

We will now assume the existence of an external weak magnetic field
${\widetilde {\bf B}}^{\rm ext}$, and see the modifications in the
chiral Lagrangian (\ref{chiralL+ e.m.}). The presence of an
external field brings a new dimensional scale in the problem. For the low energy
theory to make sense, one then has to assume that the field is
such that $p^2 \sim {\tilde e}{\widetilde
B}^{\rm ext} $.

One can get the effective field theory for the Goldstone bosons in the presence of
the magnetic field from  (\ref{chiralL+ e.m.}), simply by splitting all the vector gauge fields
into a background part and a fluctuating part
\begin{equation}
{\tilde A}_\mu \rightarrow {\tilde A}_\mu^{\rm ext} + {\tilde A}_\mu \ ,
\end{equation}
where, for example, one can take ${\tilde A}_\mu^{\rm ext} =
(0,0,x {\tilde B}^{\rm ext},0)$  to reproduce a magnetic field pointing in the $z$ direction.

With the established power counting rules and symmetry
transformation of $Q_{L/R}$, it is not difficult to see new terms
entering into the low energy Lagrangian. We will however now
neglect the effect of electromagnetic fluctuations, and consider
the direct effects of the external magnetic field, assuming
${\tilde e}^2 \ll p^2/f_\pi^2 \sim {\tilde e}{\widetilde
B}^{\rm ext} /f_\pi^2$. It is not difficult to see that a new term in the chiral Lagrangian is
allowed, namely
\begin{equation}
D ({\widetilde {\bf B}}^{\rm ext} \cdot {\widetilde {\bf B}}^{\rm ext})  {\rm Tr} \left( Q \Sigma Q \Sigma^\dagger \right)  \ .
\end{equation}

As before, this essentially represents a mass term for the charged Goldstone bosons
\begin{equation}
\label{B-chGBmass}
M^2_{\pi^\pm} = M^2_{K^\pm} = \frac{  ({\tilde e} {\widetilde B}^{\rm ext})^2 }{f^2_\pi} D  \ ,
\end{equation}
plus additional contact interactions.
Here, $D$ should be computed from the microscopic theory. It is an adimensional number, and
for the power counting to make sense, it should be of order $O(1)$.

All these considerations assume a weak magnetic field, so that the
power counting rules are valid. But one can imagine increasing the
value of the external field. As suggested by
Eq.~(\ref{B-chGBmass}), this would imply an increase of the
charged Goldstone boson masses. By strengthening the field more
and more, the charged pions and kaons  become heavier and heavier.
It will reach one point when they will become unstable. This will
happen when they have a mass of order $2 \Delta$, the energy
necessary to break a Cooper pair of quarks. If we extrapolate the
value of the mass of the charged Goldstone bosons from weak to
strong fields, this happens when
\begin{equation}
{\tilde e} {\tilde B}^{\rm ext} \sim 2 f_\pi \Delta \ ,
\end{equation}
assuming that $D$ is of order $O(1)$. To get an idea of the values
of the magnetic field we are getting, we will use the value of
$f_\pi$ of Eq.~(\ref{fpi}), for $\mu \sim 400$ MeV, and $\Delta
\sim 25$ MeV. One gets ${\tilde e} {\tilde B}^{\rm ext} \sim 5
\cdot 10^{16}$ G. For those values of the magnetic field, the
system will be formed by only five neutral Goldstone bosons. This
is precisely the number predicted from the symmetry considerations
explained in the previous subsection. We can deduce that for the
above values of the magnetic field, it will more appropriate to
say that quark matter is in the MCFL phase.

\section{Transport in the CFL phase at low temperature}
\label{transport}

 The BCS
theory explains the main phenomenological properties of electromagnetic superconductors,
such as the Meissner effect, and the appearance of an energy
gap $\Delta$ in the quasiparticle spectrum. It also explains
the reason why  superconductors are both good heat conductors and superfluids,
with exponentially suppressed transport coefficients $\sim \exp[- \Delta/T]$
for low temperatures $T \ll \Delta$.

Should we expect the transport coefficients in a
color superconductor to behave as in an electromagnetic
superconductor? Naively, one would say so, but the answer is
slightly more elaborated. Because the hydrodynamic regime of a system depends on the low energy
spectrum of the theory, a careful study of the low energy properties of every color
superconducting phase is needed before evaluating transport coefficients.

Transport properties in the CFL phase of QCD at low $T$ are not dominated by
the quarks, which certainly give a contribution of the
 sort $\sim \exp[-\Delta/T]$, as the electrons in an electromagnetic superconductor.
There are many light degrees of freedom in the CFL phase whose contribution is bigger:  nine Goldstone bosons
(or five in the presence of a strong magnetic field!),
and the ``rotated" photon. In a possible CFL quark star at finite temperature leptons are
thermally excited, so
one may also expect to find electrons. In principle, neutrinos should as well be considered
for transport. However, for the astrophysical applications we have in mind, in the temperature
regime we are going to consider all the neutrinos have already escaped from the star,
as it is deduced from their mean free path \cite{Reddy:2002xc,Jaikumar:2002vg}.

 Chiral symmetry is not an exact symmetry of QCD. Therefore, the
associated (pseudo) Goldstone bosons  are massive. Their masses
are estimated to be in the range of the tens of MeV \cite{Son:2000cm}. At sufficiently
low temperatures ($T \ll \Delta, m_\pi$) their contribution to the viscosities is Boltzmann suppressed, and
 transport is dominated by the lightest particles.
In particular, the highest contribution \cite{Shovkovy:2002kv} is given by the Goldstone
boson associated to baryon symmetry breaking, the superfluid phonon, which
remains massless. There is also a contribution of the in-medium
electromagnetism, but this turns out to be negligible.
In Ref. \cite{Manuel:2004iv}  the mean free path and shear viscosity for the superfluid phonon were computed,
finding many similarities with those of the superfluid phonon in He$^4$.
Here we will try to explain why this is so.
For all the considerations in this section, we will neglect the effects of an external magnetic field.

\subsection{Effective field theory for the superfluid phonon}

Superfluidity in He$^4$ is understood as a consequence of
Bose-Einstein condensation (BEC). The BEC and CFL superfluids have
in principle very little in common - the first system is bosonic
and non-relativistic, while the second one is fermionic and
relativistic. However, in the two systems there is a spontaneous
breaking of a global $U(1)$ symmetry, associated to particle
density in one case, to baryon charge in the other. Hence, there
is a massless collective mode or Goldstone boson in the two cases.
We will call it in the two cases superfuid phonon.\footnote{The name
of phonon was chosen by Landau, when he ``guessed"  that a
collective mode with a linear dispersion relation was needed to
explain the superfluid behavior of He$^4$. More frequently  now
``phonon" is used to denote the Goldstone boson associated to the
spontaneous breaking of  a translational symmetry, as it occurs in
crystals.}

For the two systems one can use the techniques of effective field theory to learn about the
self-interactions of the superfluid phonons. The low energy Lagrangian is  constructed using
a power expansion in momenta, taking into account properly the underlying unbroken symmetries of the system.
In a BEC the phonon $\varphi$ appears as the phase of the boson
condensate. Galilean invariance implies that the Lagrangian is
expressed in terms of powers of $(\partial_0 \varphi -
\frac{({\bf\nabla} \varphi)^2 }{2 m})$ \cite{Popov}. In the CFL system the phonon appears as a phase of
the diquark condensate. Lorentz invariance would be a symmetry of the
system if it were not for the presence of the chemical potential
$\mu$. One can treat $\mu$ as a spurion field, and then one sees
that effective Lagrangian for the phonon in this case can be
organized as a power expansion in $D_\mu \varphi= \partial_\mu
\varphi -A_\mu$, where $A_\mu = (\mu,0)$ \cite{Son:2002zn}.

The specific
coefficients of the effective Lagrangian in each case should be determined from
the microscopic theory by a matching procedure. However, this matching
actually can only be performed in the weak coupling regime of the theory.
When the system is not weakly coupled, one can still write down an effective
field theory for the low energy degrees of freedom, and expect to have
an experimental result to determine the values of those coefficients. This is the situation
that occurs for the mesons of QCD at vacuum, for example, where the value $f_\pi$ is only
determined experimentally.

In the CFL case the matching procedure can be performed for very large densities, the result is \cite{Son:2002zn}
\begin{equation}
\label{L-BGB-0}
 {\cal L}^{\rm CFL}_{\rm eff}  = \frac{3}{4 \pi^2}
\left[ (\partial_0 \varphi - \mu)^2 - ({\bf\nabla} \varphi)^2
\right]^2 + \cdots
 \end{equation}

For a BEC superfluid the effective phonon Lagrangian reads \cite{Andersen:2002nd}
\begin{equation}
\label{L-BGB-1}
 {\cal L}_{\rm eff}^{\rm BEC} = c_1
\left(\partial_0 \varphi - \frac{({\bf\nabla} \varphi)^2 }{2 m}
\right) + \frac{c_2}{2} \left(\partial_0 \varphi -
\frac{({\bf\nabla} \varphi)^2 }{2 m} \right)^2 + \cdots
\end{equation}
 For a weakly interacting bosonic system the parameters $c_1$
and $c_2$ can be computed from the microscopic theory
\cite{Andersen:2002nd}, and are given in terms of the bosonic particle
density $c_1 = n_0$, and the $s$-wave scattering length $a$, $c_2 = 1/4
\pi a$. Let us mention that a similar effective field theory, keeping not only the
Goldstone boson, but also the quantum fluctuations of the bosonic density has been
derived in Refs. \cite{Popov,Liu:1998ey}. Eq.~(\ref{L-BGB-1}) is obtained from those if one further integrates out
the density fluctuations.

 We re-scale $\varphi$ to get the standard kinetic term in both (\ref{L-BGB-0}) and (\ref{L-BGB-1}) to obtain
\begin{equation}
 \label{L-BGBcfl}
 {\cal L}_{\rm eff}^{\rm CFL}  = \frac 12 (\partial_0 \phi)^2 -
\frac{v^2}{2} ({\bf\nabla} \phi)^2 - \frac{\pi}{9 \mu^2}
\partial_0 \phi (\partial_\mu \phi  \partial^\mu \phi) +
\frac{\pi^2}{108 \mu^4}(\partial_\mu \phi  \partial^\mu
\phi)^2 \ ,
\end{equation}
and
\begin{equation}
 \label{L-BGBbec}
 {\cal L}_{\rm eff}^{\rm BEC}  = \frac 12 (\partial_0 \phi)^2 -
\frac{v^2}{2} ({\bf\nabla}  \phi)^2 - \frac{1}{2 m \sqrt{c_2}}
\partial_0 \phi ({\bf\nabla} \phi )^2 +
\frac{1}{8 m^2 c_2}({\bf\nabla}  \phi)^4 \ ,
\end{equation}
respectively. In the two cases, we have omitted a total time derivative, which is
only needed to study vortex configurations. To lowest order in momenta, the superfuid phonon dispersion relation reads
\begin{equation}
E_p = v p
\end{equation}
where, in the two cases, $v$ is given by the speed of sound of the system. More particularly,
in the CFL case $v^2 = 1/3$, while for the BEC $v^2= c_1/m c_2$. It is curious to note that even in a non-relativistic BEC the physics of the phonon is relativistic.

At low temperature the dissipation in both the CFL and BEC superfluids will be dominated by the thermal
properties of the superfluid phonons, such as damping, scattering rates, mean free paths.
These can be computed from the effective field theories displayed above.

\begin{figure}
\begin{center}
\includegraphics[width=10cm]{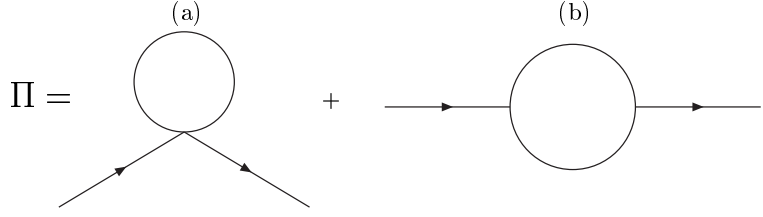}
 \caption{One-loop
contributions to the phonon self-energy.}
 \label{feyn2}
\end{center}
\end{figure}

Let us first see how the one-loop thermal physics is obtained. We will use the imaginary time
formalism (ITF) for the evaluation of the Feynman diagrams.
There are two different diagrams that contribute to
the one-loop self-energy, see Fig. \ref{feyn2}.
For external momentum $P = (p_0, {\bf p}) = (i \omega, {\bf p})$,
these are
\be
\Pi^{(a)} (P)   =
 \int \frac{d^d K}{(2 \pi)^d} \, H(P,K) S(K) \ ,
 \ee
 and
 \be
 \Pi^{(b)} (P)   =
 \int \frac{d^d K}{(2 \pi)^d} \, \Big( F(P,K) S(K) S(P-K)
\Big) \ ,
\ee
respectively, where
$S$ is the phonon propagator. In  ITF, with momentum $K= (k_0, {\bf k}) =
 ( i \omega_n , {\bf k})$, it reads
 \be
S(K) \equiv \frac{1}{\omega_n^2 + E^2_k}  \ ,
\ee
 where $\omega_n = 2 \pi nT$, with $n \in \cal Z$,
is a bosonic Matsubara frequency.
The functions $H$ and $F$ are slightly different for the CFL and BEC superfluids.
More particularly, in the first case, one has
\be
 H^{\rm CFL} (P,K)=\frac{\pi^2}{27 \mu^4} \left( 2 (K \cdot P)^2 + P^2 K^2\right) \ ,
\ee
\be
\label{cub-vertex}
 F^{\rm CFL}(P,K) =  \frac{4 \pi^2}{81 \mu^4}\left(p_0 (2
K\cdot P - K^2) + k_0 (P^2 - 2 K\cdot P) \right)^2 \ ,
\ee
while for the BEC
\be
 H^{\rm BEC}(P,K) =\frac{1}{m^2 c_2} \left( 2 ({\bf k} \cdot {\bf p})^2 + p^2 k^2\right) \ ,
\ee
\be
\label{cub-vertexBEC}
 F^{\rm BEC}(P,K) = \frac{1}{m^2 c_2} \left(p_0 (2
{\bf k\cdot p} - k^2) + k_0 (p^2 - 2 {\bf k\cdot p}) \right)^2 \ .
\ee
The BEC functions are simply obtained by replacing the four momenta by three momenta of the CFL functions, apart from an overall factor.

Since the phonon is a Goldstone boson, and  thermal effects
do not represent an explicit breaking of the $U(1)$ symmetry, its
self-energy should vanish at $P=0$. It is actually easy to check that
this holds at one-loop level
\be
 \Pi^{(a)} (P=0) =  \Pi^{(b)} (P=0) = 0 \ ,
\ee
so no thermal mass is generated. This property of the self-energy
should hold to all orders in perturbation theory.

The physics associated to the superfluid phonon is peculiar.
Even if at finite $T$ almost all particles attain a thermal mass, this
is not so for the phonon, as thermal effects do not represent
a violation of the $U(1)$ symmetry. The superfluid phonon suffers Landau damping,
which for $p_0, p \ll T$ reads
\begin{equation}
\label{Ldamp-psmall}
 {\rm Im}\, \Pi (p_0,{\bf p})  \approx    \frac{8 \pi^5}{1215}
\frac{T^4}{ v^8 \mu^4} \frac{ p^3_0} { p} \, \Theta(v^2 p^2 -
p^2_0) \ .
\end{equation}
for the CFL superfluid, while for the BEC superfluid one gets
\begin{equation}
\label{Ldamp-psmallHe4}
 {\rm Im}\, \Pi (p_0,{\bf p})  \approx  -  \frac{\pi^3}{6 m^2}
\frac{T^4}{ v^8 c_2} \frac{ p^3_0} { p} \, \Theta(v^2 p^2 -
p^2_0) \ .
\end{equation}

Evaluating this imaginary time on-shell, one finds the superfluid phonon lifetime
in the thermal bath. In the last case, one matches old results obtained using different techniques \cite{Hohenberg}.

In the same way, one can start computing the mean free path associated to different phonon-phonon
collisions. Large-angle collisions in the two cases scale with the temperature as $\sim 1/T^9$.
Small-angle collisions also scale in the two cases as $\sim 1/T^5$.
With the explanations in this Section, now one may understand the claims of Ref. \cite{Manuel:2004iv}.
 There
it was mentioned that the superfluid phonon in the CFL system had the same temperature dependence in the
damping, and also in the mean free path for small and large angle collisions that in superfluid He$^4$, in its
low temperature regime \cite{Maris}.

Still, there is a difference in the shear viscosity of the two systems, as we explain in the
following subsection.

\subsection{Shear viscosity in the phonon fluid}

Landau proposed the two fluid model of superfluidity \cite{landaufluids}, deriving the corresponding non-relativistic
hydrodynamic equations. He incorporated the effects of dissipation, taking
into account viscosity. It is noteworthy that in the superfluid there is one
shear viscosity coefficient, but three
bulk viscosities, associated to the fact that there are two fluid velocities.
The relativistic version of these hydrodynamic equations are slightly more involved.
They have been worked out in Ref. \cite{khalatnikov}.

There are several subtle points in
the computation of  transport coefficient in the low temperature regime of these
superfluids. Here we will only discuss the shear viscosity computation.

Prior to the computation of a transport coefficient, one has to evaluate the mean free path of the collisions
which are relevant for the corresponding transport phenomena.
Shear viscosity describes the relaxation of the momentum components perpendicular to the direction of transport, and it usually requires large-angle collisions. However, the computation of this transport coefficient becomes complicated by the fact that the differential
cross section of binary collisions mediated by  phonon exchange
is divergent for small-angle collisions. This is the typical
Coulomb-Rutherford collinear divergence induced by massless exchange.
In an ordinary scalar theory such a
divergence does not appear, as a thermal mass is  generated
even if the boson is massless in vacuum.
But  the phonon remains massless at finite
temperature. In Ref. \cite{Manuel:2004iv} we have suggested that
the divergence is  regulated
by the finite width of the phonon, or more precisely,
by Landau damping, a process only occurring in a thermal bath.
After regularization, we find that small-angle
processes have a shorter mean free path than large-angle ones.
This suggests that they
might be more relevant for viscosity,  as a large-angle collision
can always be achieved by the addition of many small-angle ones.

To compute the shear viscosity one solves the Boltzmann equation, linearizing it
for small departures from equilibrium. Then one finds that
zero mode occurs in the collision operator for small-angle collisions.
This happens both for the BEC superfluid \cite{Maris}, as for the
CFL superfuid \cite{Manuel:2004iv}. These zero modes act on the
direction of suppressing the small-angle collision contribution to the shear
viscosity.

In the CFL superfluid, this suppression is very severe. Could this be changed
by considering higher order effects, not contained in the lowest low energy Lagrangian
(\ref{L-BGBcfl})?
In Ref. \cite{Zarembo:2000pj}  the superfluid phonon
dispersion relation was computed to higher order, finding
\begin{equation}
\label{Zar-dr}
\omega(k) = v k \left[
1 - \frac{11}{540} \frac{k^2}{\Delta^2} + {\cal O}\left( \frac{k^4}{\Delta^4}\right)
\right] \ .
\end{equation}
As $k$ is increased, the phonons move slower, with the tendency of
suppressing collinear splitting (a phonon cannot decay into two phonons
of larger joint energy), and then the small-angle processes are kinematically
forbidden. We then found a shear viscosity coefficient \cite{Manuel:2004iv}
\begin{equation}
\label{final-all}
\eta= 1.3\cd 10^{-4} \cd \frac{\mu^8}{T^5} \ {\rm MeV}^3 \ .
\end{equation}

In superfluid He$^4$, the value of mean free path for  shear viscosity is however
compatible with that of small-angle collisions \cite{Maris}.
This led to Maris to propose a different dispersion relation for
the superfluid phonon \cite{Maris}
\begin{equation}
\omega(k) = v k \left[
1 + g(k)
\right] \ .
\end{equation}
with $g(k)$ a positive function that tends to zero for $k
\rightarrow 0$. In this way, phonons with high momenta move faster
than those of low momenta,  and can decay into
 two slower phonons. Maris proposed  different functions $g(k)$
which more or less fit the experimental value of the shear viscosity,
although the choices seem to be rather arbitrary, and not physically motivated.

Effective field theory techniques should be more appropriated to tackle the
computation of transport coefficients in a BEC at low temperature. In a weakly
coupled system, $g(k)$ could be computed from the microscopic theory to a high
level of accuracy. In a strongly coupled system, probably $g(k)$ can be determined experimentally.
 We hope to report in the future about this approach to transport phenomena in a BEC.

\section{Conclusions}

Determination of the low energy properties are essential for any study of
signatures of color superconductivity in astrophysical scenarios.

We have stressed that a strong magnetic field, such as those that
might be found in the core of some highly magnetized compact stars,
has dramatic consequences for the low energy properties of a color
superconductor, through the disappearance of low energy modes from the
spectrum. The influence of this fact on some macroscopic properties
of the superconductor has still to be analyzed.

We have presented a computation of the shear viscosity in the CFL
phase. For the analysis of r-mode instabilities bulk viscosity
coefficients are also necessary, and have not yet been computed.

For astrophysical applications, we should emphasize the following.
 At sufficiently low $T$ the phonon mean free path would exceed
the radius of a compact star. We can give a crude estimate of
the temperature when this will occur, simply by considering the equation
\be
 R < L \sim \mu^4 / T^5 \ .
\ee
If the quark chemical potential is of order $\mu \sim 500$ MeV,
and we consider $R \sim 10$ km, we find that for $T < 0.06$ MeV
superfluid phonons do not scatter within the star.
 Transport coefficients could then be dominated by the tiny contribution
of the in-medium electromagnetism, but an evaluation of the photon
mean free path also shows that for  $T \sim 0.02$ MeV it also
exceeds the radius of the star. Below that temperature,
 CFL quark matter in the star would behave as a perfect superfluid,
showing then no dissipation,
as a hydrodynamical description of the phonon and electron fluids
would be meaningless.

In a rotating superfluid there are vortices. To study the rotational
properties of a hypothetical CFL quark star, one cannot obviate that
fact. In view of our results,  the analysis of r-mode instabilities
of a CFL quark star should then be redone, taking into account both
the temperature regime of the star, and the vortex dynamics of the
CFL phase.

\acknowledgments
 I would like to thank the organizers of this
conference for giving me the opportunity to participate in this vivid meeting,
as well as my collaborators E. Ferrer, V. de la Incera, A. Dobado, F. Llanes-Estrada
and A. Rios. This work has been supported by MEC under grant No. FPA2004-00996.

\end{document}